# Competition and Price Dispersion in International Long Distance Calling

Sean Ennis[*]

US Department of Justice

October 11, 2000

## Abstract

This paper examines the relationship between changes in telecommunications provider concentration on international long distance routes and changes in prices on those routes. Overall, decreased concentration is associated with significantly lower prices to consumers of long distance services. However, the relationship between concentration and price varies according to the type of long distance plan considered. For the international flagship plans frequently selected by more price-conscious consumers of international long distance, increased competition on a route is associated with lower prices. In contrast, for the basic international plans that are the default selection for consumers, increased competition on a route is actually associated with higher prices. Thus, somewhat surprisingly, price dispersion appears to increase as competition increases.

[*]US Department of Justice, Antitrust Division, 600 E Street, NW Suite 10000, Washington, DC 20530. sfennis@aol.com. The opinions expressed here are those of the author and do not necessarily reflect those of the U.S. Department of Justice. I thank Joe Farrell, John Harkrider, Jim Lande, Bob Majure, Carl Willner and seminar participants at the FCC for their comments. Special thanks for aid with data to Linda Blake, Mark Heuritsky, Jim Lande, and Carl Willner.

# 1 Introduction

Little evidence exists about the relationship between long distance competition and long distance rates, despite the extensive and extended pursuit of such competition by regulators. In the 1990s, rate reductions in the US were dramatic while, at the same time, long distance competition increased substantially. From this association, casual empiricism might suggest that increased competition is associated with lower rates. However, such a conclusion is far too facile because the decade also witnessed a considerable decline in the costs of providing long distance service. We cannot infer that the reductions in the rates for long distance calls occurred as a result of increasing competition; instead, the rates may have fallen simply because the costs of providing service decreased. This paper seeks to better understand the relationship between long distance competition and long distance rates.

The structure of the US long distance telecommunications market has changed dramatically since MCI began providing switched long distance service in 1974. Prior to that time, AT&T was the primary provider of US long distance service. In the late 1970s, regulators encouraged competition in long distance service by prompting AT&T to negotiate temporary access tariffs that allowed long distance carriers to interconnect with its local facilities on standard terms. The fundamental rationale for policy makers to encourage competition has been the belief that competition leads to lower prices. While this belief is in accord with standard economic theory, and US long distance rates have indeed fallen by as much as 80% since 1984, there is surprisingly little evidence to support it.

The reason that price reductions could be so dramatic and yet not indicate active competition is that costs themselves have fallen substantially. Domestic US long distance rates have been affected by regulatory actions that have led to considerable reductions in the per-minute access charges that are paid by US long distance carriers to local carriers for the origination and termination of a call. Access charges have fallen from 17.26 cents per minute in 1984 to 2.85 cents per minute in 2000 (Industry Analysis Division (2000), p. 1-4, nominal values). For international long distance, the accounting rates that govern the cost of sending a call overseas from the US have also fallen considerably. Clearly, in the face of dramatic cost reductions, prices would likely decline even in the absence of competition. Thus the effect of competition on prices is not obvious despite the fact that prices have generally fallen.

The work analyzing the relationship between long distance competition and prices deals primarily with domestic US pricing (see Edelman (1997), MacAvoy (1995, 1996 and 1998), Taylor and Taylor (1993) and Taylor and Zona (1997).) One prominent strand of work, exemplified by MacAvoy (1995 and 1998) examines "basic rate" price changes between 1985 or 1987 and the early 1990s to show that prices have decreased less quickly than costs, thus indicating that margins have increased over time, while costs have fallen. This work, while certainly of value, faces two primary limitations. First of all, its measure of price typically focuses on the highest price plan (or "basic" plan.[1]) Major long distance companies have customers on a variety of different plans, and a minority of calls are actually covered by a basic rate plan. It thus ignores the majority of usage that might be indicative of actual price competition. Second, the work consists primarily

---

[1]This is the default rate received by a consumer who signs up for no special long distance plan.

of a graphical analysis of trends, and thus its statistical strength is somewhat limited.

Previous work focusing on international pricing is relatively limited (see Acton and Vogelsang (1992), Bewley and Fiebig (1988), Cave and Donnelly (1996), and Madden and Savage (2000)) and has not focused on the impact of competition, except for Madden and Savage. They analyze a limited number of countries and focus primarily on the thesis that prices fall as markets become more symmetric in their levels of competition and private ownership.[2] Madden and Savage examine only one price for each market and thus do not analyze the plan-specific pricing of carriers. The current research is distinct in that the interpretation of competitiveness variables will be clearer and, more importantly for the current project, price dispersion of different plans will be analyzed.

In this paper, we evaluate the impact of long distance competition on prices in the US by focusing on international calls originating in the US. One of the two primary questions of this paper is whether prices decrease the most when concentration decreases the most. If so, then we cannot reject the hypothesis that increased competition on routes is associated with lower prices on those routes. In particular, we examine how price changes varied depending on the change in concentration on country-pair routes between 1994 and 1998. This analysis allows for a statistical power that is absent from studies relying solely on changes in domestic interstate rates. There are more than 100 countries with annual data on prices, costs and other variables. Moreover, over the relevant time period the number of competitors on many routes has increased considerably, but to different degrees in different countries

---

[2]The interpretation of their competitiveness variables is somewhat unclear since they are all multiplied by the number of minutes of traffic even though the dependent variable of price is apparently not quantity-adjusted.

at different times.

Perhaps one of the most interesting features of the pricing of long distance calls in general, and of international long distance in particular, is that there is a wide dispersion in available prices between the low-priced flagship plans and higher-priced basic plans. Many US customers do not enroll in their primary interexchange carrier's cheapest plan for their calling patterns despite the fact that the basic plan calls can easily cost as much as ten times more than flagship plan calls. This dramatic price dispersion gives rise to the second primary question of this paper: Does price dispersion increase or decrease as competition increases? Intuitively, one might imagine that price dispersion is most likely to survive under monopolistic industry structure, but less likely to survive as the structure grows more competitive, since a competitor will likely have the incentive to offer a plan type that is, in some sense, intermediate between those of its competitors when its competitors attempt to segment the market into two or more customer types. An increasing body of literature suggests that markets subject to competition can contain significant price dispersion for reasons unrelated to cost differences (see, for example, Borenstein (1985), Shepard (1991), Borenstein (1991), Borenstein and Rose (1994), and Sorenson (2000).) These papers analyze products that are relatively, but not perfectly, homogeneous and that contain multiple sellers, such as gasoline and air travel. A common theory underlying the work is that consumer search costs explain the price dispersion. As the expected gains from search grow, price dispersion will fall. In line with this theory, Sorenson (2000) finds that repeatedly purchased pharmaceutical prescriptions, for which one would expect the greatest benefit from search, have significantly lower price dispersion than other types of prescriptions. Most surprisingly, Borenstein and Rose (1994) note that there is a significant positive effect of competition on price dispersion. That is, on a more competitive airline route, prices on the route tend to exhibit more dispersion. This result is particularly interesting because it suggests that price dispersion may not only exist in competitive markets but rather may on occasion be greater in competitive markets than in more monopolistic ones. The result is based on a cross-sectional analysis of airline fares in the second quarter of 1992.

The current data is well constructed for examining the question of whether, over time, price dispersion increases as competition increases in international long distance. In contrast to previous work, this analysis takes advantage of a panel formulation of the relationship between price dispersion and concentration, thus avoiding certain endogeneity problems that may perturb a pure cross-sectional approach. In addition, the telecommunications market benefits from having a relatively simple measure of dispersion, namely the difference between flagship and basic rates, since there are two main international long distance prices offered to residential customers by a typical long distance carrier. As a result, this analysis can provide a relatively crisp characterization of the relationship between dispersion and competition.

Variable costs can be measured with a relatively high degree of precision and thus their effect on price can be reasonably disentangled from the effect of concentration. Between 1994 and 1998, the long distance companies' costs of making international long distance calls have fallen dramatically in measurable ways. The cost to a long distance company of completing a call to a foreign country is made up of three primary parts: a local access charge paid to the local US telephone company, the network cost of bringing the call to another country, and a per-minute settlement rate that is paid to the foreign carrier that completes the call. The settlement rate has typically been negotiated by AT&T and foreign carriers under a regulatory framework that governs the allocation of return calls between carriers and the rate paid and received by a provider for sending and receiving calls. The rate negotiated by AT&T is then applied to all US carriers. From the perspective of a carrier besides AT&T, these rate changes can be viewed as exogenous cost shocks. Generally, the settlement charges constitute the vast majority of the cost of completing a call overseas.

The settlement mechanism includes a rule that when traffic is sent to the US from a foreign carrier, the foreign carrier pays a per minute settlement rate to the US carrier that is equivalent to the rate the US carrier pays for sending a call to the foreign carrier. In addition, the foreign carrier must return traffic to US carriers in proportion to the number of minutes sent to that carrier's country by the US carrier. For example, if Sprint sends 24% of the traffic from the US to a foreign country, that foreign country's carrier must return 24% of its US-bound traffic through Sprint. While a long distance carrier incurs charges for completing a call to a foreign country, it generates counterbalancing revenue when it receives traffic from that foreign carrier. Thus, from the US long distance carrier's perspective, the net cost of making calls overseas includes the settlement rates from both outgoing and incoming calls. Since the number of outgoing minutes from the US generally exceeds the number of incoming minutes, the US carriers generally face a positive per-minute cost per call.[3]

---

[3]This unusual mechanism of cost imposition was created to counterbalance the possibility that foreign carriers would charge high rates for traffic to their countries, and then negotiate low rates for return traffic with just one of the competing US carriers, leading to a higher telecom trade imbalance than already exists between the US and foreign

In the late 1990s, the rules governing the international settlement process have changed to reflect the fact that in some countries there are now competing long distance carriers. In fact, rates to some countries (e.g., Canada) can be lower than many domestic long-distance rates in the US. The Federal Communications Commission (FCC) has recently set target settlement rates for different countries which, when achieved, allow non-dominant carriers in those countries to opt out of the international settlement process. In addition, the FCC introduced in 1994 International Simple Resale (ISR) for countries that met certain competitiveness criteria within the country and which had settlement rates below particular target rates. ISR service is the provision of international switched traffic services over international private lines. ISR service allows the US-originating carrier and end-country receiving carrier to avoid the traditional settlements system.

There were significant changes in the international environment over the time period of analysis, in particular with the formation of the World Trade Organization (WTO) in 1995 and with the full or partial implementation of the WTO telecommunications agreement in selected WTO countries in 1998. Clearly it is of policy interest to see whether there have been price impacts of either WTO membership or agreements, so variables reflecting these changes will also be considered.

The rest of this paper is organized as follows: Section 2 discusses the competitive model and the method of estimating costs. Section 3 discusses the data that is used in this paper and presents descriptive statistics. Section 4 presents the empirical results. Section 5 then concludes.

---

countries (the trade imbalance for IMTS calls was about $4.8 billion in 1998 (Industry Analysis Division, 2000.) For a further discussion of the International Settlement Process, its distortions, and a proposal for a less distorted process, see Malueg and Schwartz (1998).

## 2  Model

The first question considered here is whether increasing competition on a route is associated with reductions in prices on that route. The second question considered is whether price dispersion increases or decreases as competition increases. These questions implicitly assume that pricing for different country pairs is largely governed by country-specific conditions. Given that international pricing plans typically involve prices for all countries, it is not obvious that separating prices on a route-by-route basis is appropriate. If we find that concentration levels on a route are related to prices on that route, then the fact that prices for international calling are typically bundled together does not imply that international calls are in fact a bundled product. Particularly given that much international calling is generated by foreign-born residents calling their relatives and friends in the nation of their birth, we might expect that demand for an international plan, with its bundle of prices, is frequently governed by considerations of solely one of those prices.

This work mitigates the standard endogeneity criticism of cross-sectional price-concentration studies by reporting results from the analysis of a panel data set with fixed effects. Moreover, since this study focuses on the actions of common firms in one industry facing common, route-specific cost shocks, the panel data set approach holds considerable promise. Costs are known with great precision because the vast majority of international long distance costs arise from observable, regulated settlement rates.

Before examining the relationship between prices and concentration, we should also consider how prices will change when costs are also changing. That is, we might expect that, assuming constant costs, if competition increases (or concentration decreases), industry-wide prices will fall. For simplicity, we assume that this period is exemplified primarily by changes in supply conditions, as the average cost of calls fell by about 50%. This assumption is most reasonable when the time period of analysis is short, the number of foreign-born residents is little changed, and the amount of US trade with a foreign country is litte changed.[4]

Our model relates prices over time to concentration levels over time, as well as to cost and regulatory variables. We will estimate this model in a panel framework with fixed effects for countries. That is,

$$p_{it} = D_i + h_{it} + c_{it} + WTO_{it} + ISR_{it} + e_{it}$$

where

$p_{it}$: price in country $i$ in period $t$

$D_i$: dummy set to 1 for country $i$

$h_{it}$: concentration in country $i$ in period $t$

$c_{it}$: cost in country $i$ in period $t$

$WTO_{it}$: WTO dummy variables in country $i$ in period $t$

$ISR_{it}$: ISR status dummy variable in country $i$ in period $t$

The costs include the payment made by the U.S. carrier to overseas carriers for completing calls, the payments made to U.S. local carriers for originating or completing calls, and the actual physical cost to the carrier of carrying the call.

The model will be estimated for overall average prices by all US carriers, for provider-specific average prices, and for provider-specific plan prices.

---

[4]In order to maintain our assumption that changes found in this data are primarily related to supply conditions, countries where we might expect that demand conditions have changed considerably over the time period of analysis are excluded from the data set. These include Vietnam and Eastern European countries. Their exclusion, however, does not affect the results.

In order to investigate the relationship between price dispersion and competition, we will estimate a second model that relates the degree of price dispersion to factors including the change in concentration. The model is:

$$pd_{it} = D_i + h_{it} + c_{it} + WTO_{it} + ISR_{it} + e_{it}$$

where

$pd_{it}$: difference between basic rate in country $i$ in period $t$ for a given carrier and flagship rate in country $i$ in period $t$.

We will estimate this second model in a panel framework with fixed effects for countries focusing on the price differences between basic and flagship rates, on a route by route basis. We will estimate the model for Sprint and MCI. AT&T is excluded because the framework of its flagship plan changed significantly over the period of analysis.

## 3    Data

In order to estimate the equations above, we use measures of price, cost, concentration, the WTO status of countries, and the FCC-determined ISR status of countries. We measure price as either an average price or a plan-specific price. Average price to a country is calculated by taking the domestic revenue from all international calling to a country and dividing by the number of minutes of calling to that country. Plan-specific prices detail the flagship rates of a carrier to a specific country or the basic rates charged by a carrier when consumers call a specific country. We rely on data collected by the US federal telecommunications regulator from 1994 through 1998 to estimate average prices, costs, and concentration on a route by route basis.

The average price measures are calculated first for all carriers providing international telephone service and then for the three primary individual carriers over this time period, MCI, Sprint, and AT&T. International flagship rates are calculated based on the lowest marginal rate available for calls to a foreign country.[5] These data are submitted to the FCC in regular tariff filings by each carrier. The flagship rate is a per-minute rate. Basic rates are calculated as the rate charged to a customer for calls made to foreign countries when not signed up for any special foreign rate plans. Basic rates are typically 2-3 times higher than flagship rates to a country. All price measures are adjusted for inflation with the CPI-U index from the Bureau of Labor Statistics (Bureau of Labor Statistics, 2000.) They are then logged.

The primary data source for cost and concentration consists of the FCC international data gathered according to section 43.61 of FCC regulations (FCC 1995, 1996, 1997, 1998, 1999 and 2000). Between 1994 and 1998, all carriers had to report figures related to traffic carried over their international facilities, including revenues, payments, and outgoing and incoming minutes on a route by route basis.

In order to calculate the costs of sending traffic to a given country, we begin with the net revenue paid by U.S. carriers to overseas carriers from that country and from the revenues of their return traffic from that country. From this, the cost of originating, terminating, and carrying calls to the international meeting point is subtracted to calculate the prices for Message Telephone Service (MTS) traffic to specific countries.

---

[5]Over this time period, MCI's flagship plan involved a $3.00 monthly charge and then a marginal rate per minute of usage, as did Sprint's flagship plan. The beginning of the time period is selected because 1994 is the first year for which the price data was easily available. The marginal rate is used here as the price indicator. For years when there was time-of-day pricing, the cheapest time period was selected. Adjustments to these prices for the monthly fee would be arbitrary and, to the extent that customers are selecting between plans with the same monthly fee, irrelevant, because their choice between long distance calling plans would then be influenced by the marginal rates.

The settlement costs between a carrier and the carriers of a foreign country are determined by a formula that returns traffic to the US-based carrier in proportion to the number of minutes that it sends of the total minutes sent from the US. Thus revenues derived from incoming minutes counterbalance the costs of sending outgoing minutes. Represent the total cost to carrier $i$ of transmitting $O_i$ outgoing minutes to a country as $C_i$. Then

$$C_i = s \left( O_i - \frac{O_i}{O} I \right)$$

where

$O_i$ = the number of outgoing minutes of carrier $i$

$O$ = the total number of outgoing minutes from the US

$I$ = the total number of incoming minutes to the US

$s$ = the settlement rate

The cost for originating and terminating access is derived from the FCC's table of originating and terminating charges, multiplied by the number of outgoing minutes (for the originating charge) and incoming minutes for the terminating charge. In addition, transport costs are estimated as 1 cent per minute, falling to 0.5 cents per minute by 1998. These estimates are intended to capture a known trend whose impact on route costs varied by route according to the relative traffic ratio between the US and the country at the other end of the route. This transport cost decline assumption does not affect the results. All financial variables are adjusted for inflation with the CPI-U and then logged.[6]

---

[6]There are inaccuracies in some of the data, suggesting, for instance, that rates to particular countries or territories may fluctuate by a factor of 20 from one year to the next

Concentration measures are calculated using the Hirschman-Herfindahl Index (HHI) and are based on minutes of traffic, as opposed to firm revenues and thus limit the direct impact of price on HHI. This means the HHI is not calculated from revenue information. The HHI provides a good measure of facilities-based concentration but provides an imperfect view of firm shares in the end-consumer marketplace because the FCC reports are solely for facilities-based providers. Thus these reports exclude resellers who might sell minutes to end-consumers. Thus, the HHI statistics provide a better measure of capacity concentration than retail concentration. In measuring potential competition between different facilities-based providers, focusing on capacity concentration may be preferable. Importantly, HHI values have exhibited substantial variation between 1994 and 1998. These values are shown in Table 1.

In order to reduce endogeneity in the relationship between price and concentration (which need not be severe in cases where the price considered in the regression is from a plan selected only by a small portion of all customers, and thus possibly not influencing the HHI very much at all, given that it is likely determined by the distribution of prices available), an instrumental variable for HHI is calculated as a function of the prior period's HHI for a given route and of the average current-period HHI for other countries.

In addition, we include data on the competitiveness of the destination country's telecommunications market (as indicated by if and

---

and then back by a roughly inverse factor in the following year. Such a pattern certainly suggests misreporting. These inaccuracies occur almost exclusively with respect to the countries accounting for the smallest amounts of traffic. Such inaccuracies are inherently likely in cases when there are multiple reporters who generate reports specifically for a regulatory purpose and not for an ongoing business purpose. Consequently, results will be reported for only the 100 largest revenue countries. Limiting the number of countries considered does not alter the nature or significance of the results.

when the FCC grants ISR status), on the impact of international WTO membership of the destination country, and on the impact of implementing the main WTO telecom provisions.[7] Negotiations for the telecommunications agreement were completed in 1997 and implemented in 1998 by a relatively small set of WTO members. The ISR, WTO, and WTO telecom agreement information is represented by dummy variables that are 1 in the period of change and thereafter. For instance, if a country were among the original WTO signatories, WTO membership would be indicated by a dummy set to 1 in 1995, when the WTO agreement went into effect.

Summary statistics are provided in Table 2.

## 4 Results

The results for these fixed effect regressions are reported in Tables 3-12. When considering average prices across carriers or for specific carriers, concentration has a positive and significant relationship to price, suggesting that reductions in concentration are associated with reductions in price. These results are consistent with the view that telecommunications competition leads to lower prices for consumers, even after adjusting for changes in the costs of making telephone calls.

Surprisingly, carriers appear to adjust their prices differently when setting their flagship rates and their basic rates. Lower flagship rates are associated with lower levels of concentration. However, basic rates are inversely related to levels of concentration. That is, as competition on a route increases, basic rates on that route typically increase.

---

[7]Traffic that is carried under the ISR system is still reportable under section 43.61 of FCC regulations. Thus the data relied on for much of this analysis should reflect a complete record of reported minutes and accounting payments to foreign countries over the time period in question.

These results are confirmed further by the regressions reported in Tables 11 and 12, which examine the extent of price dispersion between flagship and basic plans for MCI and for Sprint. These findings suggest that price dispersion *increases* as competition on a route increases. The relationship between concentration and price dispersion is statistically significant both for MCI and for Sprint. This result is somewhat surprising, given that one might intuitively expect that price dispersion would be largest under the most monopolistic route structures.

The finding is consistent with a carrier setting lower rates on its plan designed for the most price-elastic consumers as competition increases on a route, but at the same time raising the rates for the most inelastic consumers. The relationship between price and concentration could arise because increased competition at the low rate side of the spectrum, due to pre-paid cards, leads to lower flagship rates. As flagship rates fall, the expected benefits of search increase, leading the more price-sensitive basic rate consumers to leave for better plans. As the remaining group of basic rate consumers is more inelastic than before, it is most profitable to actually raise prices to that group at a time when concentration is generally decreasing. This suggests that at least one significant and relatively discrete set of consumers may suffer from increased levels of competition in international long distance.

Given the inverse movement of prices on the basic rate and flagship plans, it is important to consider whether, overall, increased competition hurts or helps consumers. Data limitations prevent us from knowing the quantity of minutes provided under the basic rates and under flagship plan rates. Thus, the simplest approach to answering the question of the overall impact of competition on consumers may be to examine the impact of changes in concentration on the average prices charged by a carrier. The fact that the lower average prices from providers are associated with lower concentration levels suggests that, on net, the negative impact on the basic rate consumers is outweighed by the beneficial impact on other consumers from decreased levels of concentration.

Costs, interestingly, also enter the basic rate regressions with a negative sign but a small coefficient (generally not significantly different from 0.) This may seem odd, but it is actually the case that, in a period of declining costs, the average basic rates rose substantially, from $0.97 per minute to $1.41 between 1994 and 1998 (see table 2, both for MCI and Sprint, nominal values.)

Apart from cost and US-based concentration relationships, it is important to consider the impacts of regulatory variables. The interpretation of results relating to the regulatory variables must be cautious due to the difficulty of interpreting the meaning of various regulatory and membership decisions. ISR designations are sometimes, though not systematically, associated with significantly lower prices. This may arise from the fact that ISR designations indicate a higher degree of competition in destination countries.

WTO membership appears to be associated with lower prices for international traffic over and above any effects that may arise from reduced accounting rates as a result of WTO membership, since accounting rates are included within the cost measure. This may reflect a general pro-competition bias on the part of governments that are WTO members relative to those that are not.

Countries' commitments to open their markets in accord with WTO telecom agreement principles do appear significantly related to lower prices in flagship plan rates, but not for overall average prices or for basic rates.

# 5 Conclusion

The results strongly suggest that, even for the international flagship plans offered by the major carriers, international pricing varies on a country by country basis in a way that reflects the costs of sending traffic to a given country and the level of competition to that country. This finding is interesting because generally, when consumers purchase international flagship plans, they sign up for a bundle of rates to all foreign countries. Thus one might consider that international traffic should be analyzed as a bundled product. However, the pricing relationships found here suggest that it is appropriate to consider each international route individually when evaluating competition. This is presumably because, while consumers face bundled prices, most people who make frequent international calls direct them primarily to one country, and thus care primarily about just one of the prices in the bundle when selecting their bundle.

Increased competition is associated with lower average prices for international long distance and with lower prices in flagship plans. Curiously, though, increased competition is associated with higher prices in basic plans.

Pricing displays high and increasing price dispersion. As concentration decreases, the difference increases between flagship plan prices and basic plan prices.[8] The fact that prices can move in opposite directions in plans offered by the same carrier suggests the complex nature of the demand for telecommunications services. More generally, however, this increased dispersion represents an interesting example of how offering a better deal for one plan, in response to competition, may shift the distribution of customers between the offered plans, so that raising prices for the other plan becomes profitable, as the consumers remaining in the other plan are, on average, less price-sensitive than before.

---

[8]This result helps to explain MacAvoy's finding of basic rate increases during periods of cost and concentration reductions, while at the same time suggesting that, on average, lower levels of concentration are indeed associated with lower rates.

International long distance data holds great potential for measuring the relationship between telecommunications competition and prices. The reason is that international data allows for a panel data set analysis that involves far more variation than any purely domestic analysis could provide. The findings here suggest that telecommunications competition is highly beneficial to consumers. An increase from 3 equal-sized firms to 4, for instance, may be associated with a price decrease on the order of 11.7% when considering the most important markets for telecommunications providers. While these findings relate strictly to international telephone calling, they actually reflect competition between domestic US carriers to carry domestically-originated international calls and thus can be viewed broadly as findings about domestic US competition. Overall, these results suggest that pro-competitive policies may have beneficial impacts, whether implemented by domestic policy makers or by international organizations such as the WTO.

Further theoretical research is needed to better understand the process by which increasing competition might be associated with greater price dispersion. Such a theory might focus on a changing distribution of consumers between plans when there is intense competition to provide one type of plan and limited competition to provide another. Further empirical work might focus on the increasingly significant long distance competition in countries besides the US that have opened up their markets to long distance competition. These extensions are beyond the scope of the current work, however, which is focused purely on improving our understanding of US-originated long distance calling.

The findings of this paper are consistent with the idea that increasing competition yields beneficial results for the average consumer. These findings do not necessarily mean, however, that all consumers benefit from competition. Indeed, in this instance, competition may be associated with diverging effects, depending on the degree of the consumer's price-sensitivity, as indicated by the plan chosen by the consumers. The price-sensitive consumers appear to benefit the most from competition, and this benefit appears to outweigh the harm to the less price-sensitive consumers.

Table 1: Concentration measures for International Long Distance Minutes to the Top 30 Countries

| Country | 1994 HHI | 1998 HHI | Change |
|---|---|---|---|
| Mexico | 4,769 | 3,594 | -1,175 |
| Canada | 4,013 | 4,028 | 15 |
| United Kingdom | 4,246 | 2,345 | -1,901 |
| Germany | 4,938 | 2,932 | -2,006 |
| Japan | 3,832 | 2,565 | -1,267 |
| Philippines | 4,478 | 2,644 | -1,834 |
| Korea, South | 4,058 | 3,192 | -866 |
| India | 4,155 | 3,251 | -904 |
| France | 3,868 | 2,821 | -1,047 |
| Dominican Republic | 3,057 | 1,839 | -1,218 |
| China | 3,746 | 2,243 | -1,502 |
| Italy | 5,053 | 3,581 | -1,472 |
| Taiwan | 3,706 | 2,835 | -871 |
| Colombia | 4,721 | 3,638 | -1,084 |
| Hong Kong | 3,537 | 2,836 | -701 |
| Israel | 3,694 | 3,220 | -474 |
| Brazil | 4,426 | 3,515 | -911 |
| Jamaica | 5,918 | 3,943 | -1,975 |
| El Salvador | 5,397 | 2,300 | -3,097 |
| Pakistan | 5,193 | 3,374 | -1,819 |
| Spain | 4,373 | 3,272 | -1,100 |
| Guatemala | 4,943 | 2,744 | -2,199 |
| Ecuador | 4,545 | 4,226 | -319 |
| Switzerland | 4,342 | 2,802 | -1,539 |
| Netherlands | 3,859 | 2,336 | -1,523 |
| Argentina | 3,985 | 2,868 | -1,118 |
| Peru | 5,084 | 3,742 | -1,342 |
| Thailand | 4,112 | 3,492 | -620 |
| Saudi Arabia | 4,397 | 3,727 | -670 |
| Venezuela | 4,508 | 3,581 | -926 |

Source: Calculations from FCC 43.61 Data

Table 2: Summary Statistics for 1994 and 1998

| Variable | Obs | Mean | Std. Dev. | Min | Max |
|---|---|---|---|---|---|
| hhi94 | 100 | 0.473 | 0.080 | 0.306 | 0.841 |
| hhi98 | 100 | 0.333 | 0.082 | 0.176 | 0.716 |
| isr94 | 100 | 0.020 | 0.141 | 0.000 | 1.000 |
| isr98 | 100 | 0.160 | 0.368 | 0.000 | 1.000 |
| wto94 | 100 | 0.000 | 0.000 | 0.000 | 0.000 |
| wto98 | 100 | 0.820 | 0.386 | 0.000 | 1.000 |
| wtotel94 | 100 | 0.000 | 0.000 | 0.000 | 0.000 |
| wtotel98 | 100 | 0.200 | 0.402 | 0.000 | 1.000 |
| pri94 | 100 | 0.806 | 0.235 | 0.240 | 2.078 |
| pri98 | 100 | 0.477 | 0.152 | 0.189 | 1.198 |
| costpm94 | 100 | 0.383 | 0.203 | 0.081 | 1.308 |
| costpm98 | 100 | 0.223 | 0.118 | 0.032 | 0.507 |
| MCI average price 94 | 100 | 0.700 | 0.216 | 0.243 | 1.690 |
| MCI average price 98 | 100 | 0.472 | 0.209 | 0.162 | 1.234 |
| MCI flagship price 94 | 82 | 0.509 | 0.160 | 0.115 | 1.119 |
| MCI flagship price 98 | 83 | 0.387 | 0.195 | 0.074 | 0.969 |
| MCI basic price 94 | 82 | 0.972 | 0.261 | 0.317 | 2.167 |
| MCI basic price 98 | 83 | 1.416 | 1.242 | 0.331 | 11.831 |
| MCI cost per minute 94 | 100 | 0.401 | 0.225 | 0.084 | 1.458 |
| MCI cost per minute 98 | 100 | 0.217 | 0.129 | 0.007 | 0.696 |
| Sprint average price 94 | 100 | 0.757 | 0.245 | 0.230 | 1.890 |
| Sprint average price 98 | 100 | 0.467 | 0.241 | 0.124 | 1.839 |
| Sprint flagship price 94 | 82 | 0.590 | 0.201 | 0.128 | 1.283 |
| Sprint flagship price 98 | 83 | 0.389 | 0.186 | 0.061 | 0.938 |
| Sprint basic price 94 | 82 | 0.972 | 0.261 | 0.317 | 2.167 |
| Sprint basic price 98 | 83 | 1.416 | 1.242 | 0.331 | 11.831 |
| Sprint cost per minute 94 | 100 | 0.393 | 0.195 | 0.030 | 1.027 |
| Sprint cost per minute 98 | 100 | 0.221 | 0.174 | -0.268 | 1.244 |
| ATT price 94 | 100 | 0.870 | 0.244 | 0.255 | 2.203 |
| ATT price 98 | 100 | 0.617 | 0.222 | 0.193 | 1.727 |
| ATT cost per minute 94 | 100 | 0.373 | 0.207 | 0.076 | 1.359 |
| ATT cost per minute 98 | 100 | 0.250 | 0.149 | 0.041 | 0.906 |

Dollar figures are nominal and not logged.

Table 3: All International Long Distance Average Price: 1994-1998

Dependent variable: average price of all long distance carriers

| | |
|---|---|
| hhi | 1.821* |
| | (8.262) |
| costpm | 0.277* |
| | (5.922) |
| isr | -0.121 |
| | (-1.594) |
| wto | -0.030 |
| | (-1.077) |
| wtotel | -0.098 |
| | (-1.501) |
| n | 500 |
| groups | 100 |
| $R^2$ within | 0.4722 |

Average price to a country is calculated by taking the domestic revenue from all international calling to a country and dividing by the number of minutes of calling to that country.

Regression with fixed country effects. t-statistics in parentheses.

*Significant at 0.01 level

**Significant at 0.05 level

***Significant at 0.10 level

Table 4: MCI International Long Distance Average Price: 1994-1998

Dependent variable: MCI average price across all plans

| | | |
|---|---|---|
| hhi | 1.970* | 1.989* |
| | (7.754) | (7.895) |
| costpm | 0.041 | - |
| | (0.768) | |
| mcostpm | - | 0.023 |
| | | (0.638) |
| isr | -0.321* | -0.329* |
| | (-3.659) | (-3.811) |
| wto | -0.014 | -0.016 |
| | (-0.433) | (-0.529) |
| wtotel | -0.074 | -0.077 |
| | (-0.985) | (-1.036) |
| n | 500 | 500 |
| groups | 100 | 100 |
| $R^2$ within | 0.314 | 0.313 |

Average price to a country is calculated by taking the MCI domestic revenue from all international calling to a country and dividing by the number of MCI minutes of calling to that country.

Regression with fixed country effects. t-statistics in parentheses.

*Significant at 0.01 level

**Significant at 0.05 level

***Significant at 0.10 level

Table 5: MCI International Long Distance Flagship Plans: 1994-1998

Dependent variable: MCI flagship plan rates

|  |  |  |
|---|---|---|
| hhi | 1.137* | 1.700* |
|  | (5.151) | (7.407) |
| costpm | 0.381* | - |
|  | (8.406) |  |
| mcostpm | - | 0.110* |
|  |  | (3.612) |
| isr | -0.114*** | -0.204* |
|  | (-1.650) | (-2.776) |
| wto | 0.079* | 0.049*** |
|  | (3.029) | (1.769) |
| wtotel | -0.226* | -0.285* |
|  | (-3.853) | (-4.524) |
| n | 414 | 414 |
| groups | 83 | 83 |
| $R^2$ within | 0.5358 | 0.457 |

Regression with fixed country effects. t-statistics in parentheses.
*Significant at 0.01 level
**Significant at 0.05 level
***Significant at 0.10 level

Table 6: MCI International Long Distance Basic Rates: 1994-1998

Dependent variable: MCI basic rates

| | | |
|---|---|---|
| hhi     | -0.748*  | -0.756*  |
|         | (-4.004) | (-4.211) |
| costpm  | -0.048   | -        |
|         | (-1.244) |          |
| mcostpm | -        | -0.034*  |
|         |          | (-1.404) |
| isr     | -0.054   | -0.049   |
|         | (-0.933) | (-0.852) |
| wto     | 0.107*   | 0.110*   |
|         | (4.852)  | (5.051)  |
| wtotel  | 0.015    | 0.020    |
|         | (0.298)  | (0.399)  |
| n       | 414      | 414      |
| groups  | 83       | 83       |
| $R^2$ within | 0.2482 | 0.2491 |

Regression with fixed country effects. t-statistics in parentheses.
*Significant at 0.01 level
**Significant at 0.05 level
***Significant at 0.10 level

Table 7: Sprint International Long Distance Average Price: 1994-1998

Dependent variable: Sprint average price across all plans

| | | |
|---|---|---|
| hhi | 1.714* | 2.181* |
| | (6.602) | (8.528) |
| costpm | 0.309* | - |
| | (5.616) | |
| scostpm | - | 0.096* |
| | | (2.622) |
| isr | -0.108 | -0.182** |
| | (-1.201) | (-2.00) |
| wto | -0.144* | -0.161* |
| | (-4.459) | (-4.815) |
| wtotel | -0.020 | -0.057 |
| | (-0.256) | (-0.727) |
| n | 499 | 495 |
| groups | 100 | 100 |
| $R^2$ within | 0.446 | 0.413 |

Average price to a country is calculated by taking the Sprint domestic revenue from all international calling to a country and dividing by the number of Sprint minutes of calling to that country.

Regression with fixed country effects. t-statistics in parentheses.

*Significant at 0.01 level
**Significant at 0.05 level
***Significant at 0.10 level

Table 8: Sprint International Long Distance Flagship Plans: 1994-1998

Dependent variable: Sprint flagship plan rates

| | | |
|---|---|---|
| hhi | 1.102* | 1.418* |
| | (5.389) | (7.093) |
| costpm | 0.314* | - |
| | (7.479) | |
| scostpm | - | 0.149* |
| | | (4.998) |
| isr | -0.002 | -0.078 |
| | (-0.037) | (-1.203) |
| wto | -0.081* | -0.099* |
| | (-3.368) | (-3.990) |
| wtotel | -0.224* | -0.241* |
| | (-4.107) | (-4.286) |
| n | 414 | 411 |
| groups | 83 | 83 |
| $R^2$ within | 0.5695 | 0.5353 |

Regression with fixed country effects. t-statistics in parentheses.
*Significant at 0.01 level
**Significant at 0.05 level
***Significant at 0.10 level

Table 9: Sprint International Long Distance Basic Rates: 1994-1998

Dependent variable: Sprint basic rates

| | | |
|---|---|---|
| hhi | -0.748* | -0.817* |
| | (-4.004) | (-4.580) |
| costpm | -0.048 | - |
| | (-1.244) | |
| scostpm | - | -0.013 |
| | | (-0.475) |
| isr | -0.054 | -0.041 |
| | (-0.933) | (-0.719) |
| wto | 0.107* | 0.112* |
| | (4.852) | (5.065) |
| wtotel | 0.015* | 0.021 |
| | (0.298) | (0.411) |
| n | 414 | 411 |
| groups | 83 | 83 |
| $R^2$ within | 0.2482 | 0.2456 |

Regression with fixed country effects. t-statistics in parentheses.
*Significant at 0.01 level
**Significant at 0.05 level
***Significant at 0.10 level

Table 10: AT&T International Long Distance: 1994-1998

Dependent variable: AT&T average price across all plans

| | | |
|---|---|---|
| hhi | 1.069* | 1.591* |
| | (3.694) | (5.897) |
| costpm | 0.164* | - |
| | (2.678) | |
| acostpm | - | -0.109** |
| | | (1.819) |
| isr | -0.143 | -0.237* |
| | (-1.435) | (-2.365) |
| wto | -0.018 | -0.043 |
| | (-0.510) | (-1.185) |
| wtotel | -0.073 | -0.117 |
| | (-0.860) | (-1.381) |
| n | 499 | 499 |
| groups | 100 | 100 |
| $R^2$ within | 0.171 | 0.163 |

Average price to a country is calculated by taking the domestic AT&T revenue from all international calling to a country and dividing by the number of AT&T minutes of calling to that country.

Regression with fixed country effects. t-statistics in parentheses.
*Significant at 0.01 level
**Significant at 0.05 level
***Significant at 0.10 level

Table 11: MCI International Long Distance: Basic-Flagship Price Dispersion 1994-1998

Dependent variable: MCI price dispersion

| | | |
|---|---|---|
| hhi      | -1.885*  | -2.456*  |
|          | (-6.843) | (-8.742) |
| costpm   | -0.429*  | -        |
|          | (7.578)  |          |
| mcostpm  | -        | -0.143*  |
|          |          | (-3.848) |
| isr      | 0.059    | 0.155*   |
|          | (0.689)  | (1.723)  |
| wto      | 0.028    | 0.061    |
|          | (0.864)  | (1.783)  |
| wtotel   | 0.241*   | 0.305*   |
|          | (3.289)  | (3.951)  |
| n        | 414      | 414      |
| groups   | 83       | 83       |
| $R^2$ within | 0.570 | 0.517   |

Price dispersion for a given year is calculated by taking the basic rate charged by MCI for international calling to a country and subtracting MCI's flagship plan rate to that country.

Regression with fixed country effects. t-statistics in parentheses.

*Significant at 0.01 level
**Significant at 0.05 level
***Significant at 0.10 level

Table 12: Sprint International Long Distance: Basic-Flagship Price Dispersion 1994-1998

Dependent variable: Sprint price dispersion

|  |  |  |
|---|---|---|
| hhi | -1.850* | -2.235* |
|  | (-6.485) | (-8.021) |
| costpm | -0.362* | - |
|  | (-6.175) |  |
| scostpm | - | -0.161* |
|  |  | (-3.890) |
| isr | -0.052 | -0.036 |
|  | (-0.584) | (-0.403) |
| wto | 0.187* | 0.210* |
|  | (5.592) | (6.106) |
| wtotel | 0.238* | 0.261* |
|  | (3.139) | (3.337) |
| n | 414 | 414 |
| groups | 83 | 83 |
| $R^2$ within | 0.575 | 0.548 |

Price dispersion for a given year is calculated by taking the basic rate charged by Sprint for international calling to a country and subtracting Sprint's flagship plan rate to that country.

Regression with fixed country effects. t-statistics in parentheses.

*Significant at 0.01 level
**Significant at 0.05 level
***Significant at 0.10 level